\documentclass[conference]{IEEEtran}
\IEEEoverridecommandlockouts
\usepackage{amsmath,amssymb,amsfonts}
\usepackage{algorithmic}
\usepackage{graphicx}
\usepackage{textcomp}
\def\BibTeX{{\rm B\kern-.05em{\sc i\kern-.025em b}\kern-.08em
    T\kern-.1667em\lower.7ex\hbox{E}\kern-.125emX}}

\usepackage[utf8]{inputenc}
\usepackage[english]{babel}
\usepackage{fancyhdr}

\usepackage{amsmath} 
\usepackage{amssymb}  
\usepackage{gensymb}
\usepackage{amsfonts} 
\usepackage{bm}
\usepackage{bbm}
\usepackage{graphicx}
\usepackage[table,xcdraw,dvipsnames]{xcolor}
\usepackage{subcaption}
\usepackage[font={small}]{caption}
\usepackage{multirow}
\usepackage{pgfplots}
\pgfplotsset{compat = newest}

\usepackage{url}
\usepackage{lastpage}
\usepackage[hidelinks]{hyperref}

\usepackage[backend=bibtex,
doi=false,isbn=false,url=false,
natbib=true, sorting=none]{biblatex}
\usepackage{biblatex}
\DeclareMathAlphabet{\mymathbb}{U}{BOONDOX-ds}{m}{n}
\newcommand{\bb}[1]{\mymathbb{#1}}

\title{Maximum likelihood estimation of distribution grid topology and parameters from smart meter data
\thanks{Research supported by the Swiss National Science Foundation under the NCCR Automation (grant agreement 51NF40\_180545).}
}

\author{\IEEEauthorblockN{1\textsuperscript{st} Lisa Laurent}
\IEEEauthorblockA{\textit{Institute of Electrical Engineering}\\
\textit{EPFL}\\
Lausanne, Switzerland \\
lisa.laurent@epfl.ch}
\and
\IEEEauthorblockN{2\textsuperscript{nd} Jean-S\'ebastien Brouillon}
\IEEEauthorblockA{\textit{Institute of Mechanical Engineering} \\
\textit{EPFL}\\
Lausanne, Switzerland \\
jean-sebastien.brouillon@epfl.ch}
\and
\IEEEauthorblockN{3\textsuperscript{rd} Giancarlo Ferrari-Trecate}
\IEEEauthorblockA{\textit{Institute of Mechanical Engineering} \\
\textit{EPFL}\\
Lausanne, Switzerland\\
giancarlo.ferraritrecate@epfl.ch}
}

\begin{document}

\maketitle
\pagestyle{plain}
\newcounter{appcounter}
\setcounter{appcounter}{1}

\begin{abstract}
This paper defines a Maximum Likelihood Estimator (MLE) for the admittance matrix estimation of distribution grids, utilising voltage magnitude and power measurements collected only from common, unsychronised measuring devices (Smart Meters). First, we present a model of the grid, as well as the existing MLE based on voltage and current phasor measurements. Then, this problem formulation is adjusted for phase-less measurements using common assumptions. The effect of these assumptions is compared to the initial problem in various scenarios. Finally, numerical experiments on a popular IEEE benchmark network indicate promising results. Missing data can greatly disrupt estimation methods. Not measuring the voltage phase only adds 30\% of error to the admittance matrix estimate in realistic conditions. Moreover, the sensitivity to measurement noise is similar with and without the phase.
\end{abstract}

\section{INTRODUCTION}
New intermittent energy agents in distribution network e.g. wind, solar, storage and controllable loads, bring major changes into the grid operations and planning. In the past, power generation used to be centralised which implies unidirectional electrical power flows from higher to lower voltage levels. Today, grids are evolving towards decentralised energy generation and bi-directional flows. Although it offers the possibility of better optimising the power flows in the network, it requires a better knowledge of the grid topology and line parameters, which are contained in the admittance matrix. However, obtaining an exact estimation of this matrix remains difficult since the admittances can either be unavailable or deviate from their exact values due to topology changes or external factors, such as faults.


The recent installation of numerous sensors in distribution grids has developed interest for automatic, data-driven approaches to identify the admittance matrix. Most of the recent contributions require measurements from micro Phasor Measurement Units ($\mu$PMUs) \cite{tomlin, moffat_2019, wehenkel2020parameter, eiv, patopa}. However, today's grid measurements mainly come from Smart Meters, which measure the nodal active and reactive power injection as well as the nodal voltage magnitude without requiring time synchronisation. Micro-PMUs, in contrast, use a GPS signal to measure the phase angles between each nodes precisely. The aforementioned studies show that parameters identification, even with synchronised data, is a very challenging task. Moreover, an accurate measurement of voltages is crucial as their variations are very small compared to the rated value, which means that any erroneous or missing measurement can quickly make the identification problem ill-posed.

Only few papers address the grid parameters estimation problem using Smart Meters, and they are all based on Ordinary Least-Squares (OLS) regressions \cite{zhang2021distribution, cn, srinivas2022topology}. In order to refine the parameters identified with OLS, the works \cite{cn} and \cite{srinivas2022topology} use a Newton-Raphson method, which relies heavily on tuned constants and initial point, often leading to case-dependent results. Moreover, the authors of \cite{zhang2021distribution} and \cite{cn} use a noise on voltage measurements that may be highly correlated to the signal, and \cite{srinivas2022topology} does not specify how this noise is generated. Finally, \cite{srinivas2022topology} uses a mix of $\mu$PMUs and Smart Meters which is realistic at medium-high voltages but not at low voltages.

The aim of this paper is to provide a frequentist approach to the identification problem with Smart Meter data, independent from tuning parameters requiring knowledge of the system. To do so, we will adapt the Maximum Likelihood Estimator (MLE) from \cite{eiv} because it is a statistically efficient and unbiased estimator. In addition, it allows the use of prior knowledge, if present, in a mathematically supported way through a Bayesian framework. The main contribution of this paper is the reformulation of the MLE when the phase is unknown using the linearised power-flow equations.

The paper is organised as follows. First, a grid model is presented to formally define the admittance matrix, then the Maximum Likelihood Estimator problem is stated. The latter is then transformed to adapt to Smart Meter measurements. Then, the necessary approximations are discussed by comparing them theoretically with the power flow equations. Finally, a case study is conducted on the IEEE 33 bus benchmark network to verify the theoretical results and to study the effect of various noise levels.

\subsection{Preliminaries and notations}
\noindent
Let $j = \sqrt{-1}$ denote the imaginary unit. For $\underline{x} \in \mathbb{C}$, $\underline{x}^*$ is its complex conjugate. $\bm{x} \in \mathbb{R}^n$ denotes a vector, $X \in \mathbb{R}^{m\times n}$ denotes a matrix of size $m$-by-$n$ and $X^T$ is its transpose. $\mathcal{I}_{m \times n}$ is the $m$-by-$n$ identity matrix. For a $(m,n)$ matrix X, $|| X ||_F$ is its Frobenius norm defined as $|| X ||_F = \sqrt{\sum_{i=1}^m \sum_{j=1}^n |x_{ij}|^2 }$, and $\text{vec}(X)$ is its column vectorisation defined as $\text{vec}(X) = \begin{bmatrix} x_{11}, \ldots, x_{m1}, x_{12}, \ldots, x_{m2}, \ldots, x_{1n}, \ldots, x_{mn}, \end{bmatrix}^T$. 
The operator $\otimes$ refers to the Kronecker product and
the operators $\odot$ and $\oslash$ to the Hadamard (element-wise) product and division, respectively.
$\mathcal{X} \sim \mathcal{N}(\mu,\,\sigma^{2})$ denotes a Gaussian random variable with expected value $\text{E}[\mathcal{X}]=\mu$ and variance $\text{Var}[\mathcal{X}] = \sigma^2$.

\section{BACKGROUND}
\subsection{Grid model}
An electric distribution network is modelled as an undirected, weighted and connected graph, where the $n$ nodes represent buses, generators or loads and the edges are the power lines. Each edge $(h,k)$ is associated to a complex weight equal to the line admittance $\underline{y}_{hk} = g_{hk} + j b_{hk}$, where $g_{hk} > 0$ is the line conductance and $b_{hk} \in \mathbb{R}$ the line susceptance between the nodes $h$ and $k$, for $h,k=1,\ldots,n$.

A power network is completely represented by the $n$-by-$n$ complex admittance matrix $\underline{Y}$. Its off-diagonal elements are equal to the opposite of the admittance: $\underline{Y}_{hk} = - \underline{y}_{hk}$ for $h\ne k$, and its diagonal elements are equal to the sum of all the admittances of the branches connected to node $h$ including the ones with the neutral, called shunt admittances: $\underline{Y}_{hh} = \underline{y}_{ho} + \sum_{k=1, k \ne h}^{n} \underline{y}_{hk}$. 

\textbf{Assumption \theappcounter}: the admittance matrix is symmetrical.\\
\stepcounter{appcounter}
The admittance matrix is typically sparse and diagonal-dominant because each of its diagonal elements, in absolute value, is not lower than the sum of the other elements in the same row. If there is no phase-shifting transformers nor series capacitors, the matrix is symmetric.

The admittance matrix can be decomposed into the conductance and the susceptance matrices, respectively denoted by $G \in \bb R^{n \times n}$ and $B \in \bb R^{n \times n}$, such that $\underline{Y}_{hk} = G_{kh} + j B_{hk}$. 

The current-voltage relation arising from the Kirchhoff's and Ohm's laws gives: 
\begin{equation}
    \label{eq:cv}
    \underline{\bm{i}}=\underline{Y}\underline{\bm{v}},
\end{equation}
with $\underline{\bm{v}} = \bm{v} e^{j\bm{\theta}} \in \mathbb{C}^n$ and $\underline{\bm{i}} = \bm{i} e^{j\bm{\phi}} \in \mathbb{C}^n$. $\bm{\theta}$ and ${\bm{\phi}}$ are vectors of the voltage and current phases for all nodes.



\subsection{Power Flow Equations}
If the network structure and the admittances constituting it are known, and if the voltages in magnitude and phase are observed, all other electrical quantities of interest can be computed. Namely, the power and the currents injected or extracted from the nodes, the power flows and currents along the branches and the network losses. Therefore, the problem of state estimation of the grid is equivalent to determine the voltage phasors in all the network nodes.

\hypertarget{app2}{}
\textbf{Assumption \theappcounter}: the voltage angle difference $|\theta_h - \theta_k |$ is usually within 5\degree or 0.1 rad. \stepcounter{appcounter}\\
Studies \cite{voltapp} have shown that such approximation, while significantly reducing the computational time, has very little impact on the optimal solution. According to the \hyperlink{app2}{assumption 2}, the power flow equations can be linearised considering small angles limitations. The Taylor expansion is then applied around the point where all angles are the same. It yields:
\begin{subequations}
\label{eq:pf_lin}
\begin{align}
    & p_h(\bm{v}, \bm{\theta}) \approx v_h \sum_{k=1}^{n} v_k G_{hk} + v_h \sum_{k \neq h }^{} v_k B_{hk} (\theta_h - \theta_k), \\
    & q_h(\bm{v}, \bm{\theta}) \approx -v_h \sum_{k=1}^{n} v_k B_{hk} + v_h \sum_{k \neq h }^{} v_k G_{hk} (\theta_h - \theta_k).
\end{align}
\end{subequations}
This approximation has been used in other studies to estimate the voltage phase with a different approach \cite{patopa, cn}.

\section{PROBLEM STATEMENT}
The problem consists in identifying the admittance matrix using measurements of the voltage and current at different steady-states of the system.
Let N be the number of samples in time and $\underline{\bm{v}}_t \in \mathbb{C}^n$ and $\underline{\bm{i}}_t \in \mathbb{C}^n$ be the vectors of nodal voltage and current injections for time $t=1,...,N$, $n$ being the number of nodes, then equation (\ref{eq:cv}) transforms into:
\begin{equation}
\label{eq:IYV}
    \underline{I}=\underline{V}\underline{Y},
\end{equation}
with $\underline{V} = [\underline{\bm{v}}_1, \underline{\bm{v}}_2, \ldots, \underline{\bm{v}}_N]^T \in \mathbb{C}^{N\times n}$ and $\underline{I} = [\underline{\bm{i}}_1, \underline{\bm{i}}_2, \ldots, \underline{\bm{i}}_N]^T \in \mathbb{C}^{N\times n}$.
In practice, noise is added to the measurements $\underline{\tilde{V}}$ and $\underline{\tilde{I}}$ of the voltages and currents by the sensors\footnote{The measurements are centered around their time average for robustness.}:
\begin{subequations}
\label{eq:noisy_var}
\begin{align}
    & \underline{\tilde{V}} = \underline{V} + \Delta \underline{V}, \\
    & \underline{\tilde{I}} =\underline{I} + \Delta \underline{I}.
\end{align}
\end{subequations}
Inserting (\ref{eq:noisy_var}) in (\ref{eq:IYV}) gives
\begin{equation}
\label{eq:noisy_IYV}
    \underline{\tilde{I}} - \Delta \underline{I} = (\underline{\tilde{V}} - \Delta \underline{V}) \underline{Y}.
\end{equation}
In power systems, sensors often measure active and reactive powers instead of line currents. Hence, one needs to express (\ref{eq:noisy_IYV}) in terms of power and voltage rather than current. To do so, the formula of the apparent power, $ (\underline{\tilde{I}} - \Delta \underline{I}) = (\underline{S} \oslash (\underline{\tilde{V}} - \Delta \underline{V}))^*$ where $S = P + jQ$, is incorporated to (\ref{eq:noisy_IYV}) and gives the following reformulation.
\begin{equation}
\label{eq:noisy_YV}
    (\underline{S} \oslash (\underline{\tilde{V}} - \Delta \underline{V}))^* = (\underline{\tilde{V}} - \Delta \underline{V}) \underline{Y}.
\end{equation}

\section{DATA AND MEASUREMENT NOISE}
\hypertarget{sectionIV}{}
\textbf{Assumption \theappcounter}: Smart Meters providing the measurements are placed such that the grid is observable. \\
\stepcounter{appcounter}
A grid is observable if, for a given set of measurements (in terms of number, type, and location), it is possible to infer the entire admittance matrix.
The study \cite{pmugaussian} has shown that measurement errors of $\mu$PMUs and Smart Meters can be assumed to have a Gaussian distribution with zero mean and constant variance in polar coordinates. Let $\tilde{v}$ and $\tilde{\theta}$ be the measured magnitude and phase and $v$ and $\theta$ their corresponding actual values. Then $\tilde{v} = v + \epsilon$ and $\tilde{\theta} = \theta + \delta$, where $\epsilon \sim \mathcal{N}(0,\,\sigma_\epsilon^{2})$ and $\delta \sim \mathcal{N}(0,\,\sigma_\delta^{2})$. Moreover, as the samples at one instant do not use measurements of another instant, and because different sensors are placed at different nodes, one can make the following assumption.

\textbf{Assumption \theappcounter}: samples taken at different time steps on the same node and samples taken on two different nodes at the same time are independent. \\
\stepcounter{appcounter}
The noise variance can therefore be expressed for a single measurement without loss of generality.

In cartesian coordinates, such that $ \tilde{v} e^{j\tilde{\theta}} = (c + \Delta c) + j(d + \Delta d) = \tilde{c} + j \tilde{d}$, the noise arising from the linearisation of the noise in polar coordinates gives:
\begin{subequations}
\label{eq:cart_noise}
\begin{align}
    & \text{Var}[\Delta c | \tilde{v}, \tilde{\theta}] = \sigma_\epsilon \cos ^2 \tilde{\theta} + \sigma_\delta \tilde{v}^2 \sin ^2 \tilde{\theta}, \\
    & \text{Var}[\Delta d | \tilde{v}, \tilde{\theta}] = \sigma_\epsilon \sin ^2 \tilde{\theta} + \sigma_\delta \tilde{v}^2 \cos ^2 \tilde{\theta}, \\
    & \text{Cov}[\Delta c, \Delta d | \tilde{v}, \tilde{\theta}] = \sin ^2 \tilde{\theta} \cos \tilde{\theta} \left(\sigma_\epsilon - \sigma_\delta \tilde{v}^2 \right).
\end{align}
\end{subequations}
The noise on a phasor measurement, for both the current and the voltage, is therefore modelled with:
\begin{equation}
    \label{eq:dc_dd_distrib}
    \begin{bmatrix}
    \Delta c \\ \Delta d 
    \end{bmatrix} 
    \sim \mathcal{N}(\mathbb{O}_2,\,\Sigma)
\end{equation}
with $\Sigma$ is a 2-by-2 matrix composed of the elements (\ref{eq:cart_noise}). We denote the covariance matrix of a measurement at node $h$ and time $t$ by $\Sigma_{v,th}$ and $\Sigma_{i, th}$ for the voltages and currents, respectively.

\section{METHODS}
\subsection{Admittance matrix estimation}
\subsubsection{Estimation knowing the phase} 
Maximum Likelihood Estimation (MLE) consists of estimating the parameters of an assumed probability distribution, given some observed data. This is achieved by maximising a (log)-likelihood function (or equivalently minimising its opposite) so that, under the assumed statistical model, the observed data is most probable. 
Considering $\Delta \underline{V}$ and $\Delta \underline{I}$ with Gaussian distribution, as described in \hyperlink{sectionIV}{Section IV}, the MLE is a weighted Total Least Squares estimator. Its exact expression is given in \cite{eiv}
\begin{subequations}
\label{eq:mle_pb}
\begin{align}
        \min_{\underline{\hat{Y}}, \Delta \underline{V}, \Delta \underline{I}} &{- \mathcal L(\underline{\hat{Y}}, \Delta \underline{V}, \Delta \underline{I})} \\
        \text{s.t.}\;\;\;\; &(\underline{S} \oslash (\underline{\tilde{V}} - \Delta \underline{V}))^* = (\underline{\tilde{V}} - \Delta \underline{V}) \underline{\hat{Y}}, \label{eq:mle_const} 
\end{align}
\end{subequations}
where the log-likelihood is defined by
\begin{align}\label{eq:likelihood_def}
    \!\!- \mathcal L(\underline{\hat{Y}}\!, \Delta \underline{V}, \Delta \underline{I}) =\!\!
    \sum_{h=1}^n \sum_{t=1}^N 
    &\|[\Re(\Delta \underline{V}_{th}),  \Im(\Delta \underline{V}_{th})]\|_{\Sigma_{v,th}^{-1}} \!\!
    \\ \nonumber
    &+ \|[\Re(\Delta \underline{I}_{th}), \Im(\Delta \underline{I}_{th})]\|_{\Sigma_{i,th}^{-1}}.
\end{align}
The expression \eqref{eq:likelihood_def} can also be written in vector form, which enables fast computation \cite{eiv}.
 

\subsubsection{Problem adjustment to estimate the phase}
If the voltage phases are not recorded by the measuring devices, the MLE problem could be adjusted by replacing the constraint (\ref{eq:mle_const}) with the linearised power equations (\ref{eq:pf_lin}). This would however be much more computationally intensive, as the constraint (\ref{eq:mle_const}) could no longer be written in vector form, i.e. for all nodes and time steps at once. 

Alternatively, according to the \hyperlink{app2}{assumption 2}, the voltage phasors can be approximated by its magnitude:
\begin{equation}
\label{eq:volt_approx}
    (\underline{\tilde{V}} - \Delta \underline{V}) \approx (\tilde{V} - \Delta \underline{V}).
\end{equation}
Note that $\tilde{V} \in \mathbb{R}$ thus the voltage noise $\Delta \underline{V}$ captures both the real and complex error of this approximation.
The MLE problem (\ref{eq:mle_pb}) in its approximated form is given by
\begin{subequations}
\label{eq:mle_pb_app} 
\begin{align}
        \min_{\underline{\hat{Y}}, \Delta \underline{V}, \Delta \underline{I}} &{- \mathcal L(\underline{\hat{Y}}, \Delta \underline{V}, \Delta \underline{I})}\\
        \text{s.t.}\;\;\;\; &(\underline{S} \oslash (\tilde{V} - \Delta \underline{V}))^* \approx (\tilde{V} - \Delta \underline{V}) \underline{\hat{Y}} \label{eq:mle_const_app} 
\end{align}
\end{subequations}

\subsection{Error evaluation}
\subsubsection{Constraint reformulation}
To assess the effect of the approximations made to formulate the phase-less MLE problem (\ref{eq:mle_pb_app}), the constraint (\ref{eq:mle_const_app}) can be compared to the power flow equations. For that purpose, the constraint has to be manipulated to separate the apparent power into its active and reactive terms and to express them in index notation form. 

Two approximations have to be introduced. The former (\ref{eq:voltage_app}) relies on the \hyperlink{app2}{assumption 2} stating that the voltage angle $\theta$ is small enough to use the small angle approximation. The second approximation (\ref{eq:current_app}) omits the noise on the voltage measurements. Note that the authors of \cite{cn} use the same two approximations.
\begin{subequations}
\label{eq:app}
\begin{align}
    & \underline{V} \approx V + jV \theta \label{eq:voltage_app},\\
    & \underline{S} \oslash (\tilde{V} - \Delta \underline{V}) \approx \underline{S} \oslash V. \label{eq:current_app}
\end{align}
\end{subequations}

Separating the active and reactive power, the constraint (\ref{eq:mle_const_app}) of the approximated MLE problem becomes
\begin{align}
    \label{eq:const_pq}
    \begin{bmatrix}
    P \oslash V \\
    -Q \oslash V
    \end{bmatrix} \approx \begin{bmatrix}
    G & -B \\ B & G
    \end{bmatrix}
    \begin{bmatrix}
    V \\
    V \odot \theta
    \end{bmatrix}.
\end{align}
In index notation form, it corresponds for each time instant to
\begin{subequations}
\label{eq:p_approx}
\begin{align}
    & p_h(\bm{v}, \bm{\theta}) \approx v_h \! \sum_{k=1}^{n} \!v_k G_{hk} \!+\!\, v_h \! \sum_{k\neq h}^{}\! B_{hk} \!\left( v_h \theta_h \!-\! v_k \theta_k \right)\!, \\
    & q_h(\bm{v}, \bm{\theta})  \approx -v_h \! \sum_{k=1}^{n} \!v_k B_{hk} \!+\!\,v_h \! \sum_{k\neq h}^{} \! G_{hk} \!\left( v_h \theta_h \!-\! v_k \theta_k \right)\!.
\end{align}
\end{subequations}

\subsubsection{Equations comparison}
\hypertarget{subsectionVB2}{}
The difference between the equations obtained from the linearised power flow equations (\ref{eq:pf_lin}) and the ones obtained from the adapted MLE constraint (\ref{eq:p_approx}) reads as:
\begin{subequations}
\label{eq:diff_powers}
\begin{align}
    & |\Delta p_h(\bm{v}, \bm{\theta})| = v_h \theta_h \sum_{k \ne h} B_{hk} (v_h - v_k), \\
    & |\Delta q_h(\bm{v}, \bm{\theta})| = v_h \theta_h \sum_{k \ne h} G_{hk} (v_h - v_k).
\end{align}
\end{subequations}

The voltage angle $\theta \ll 1$ and magnitude differences $v_h - v_k \ll 1$ being small, their product results in a term of second order. As a consequence, the term of difference \eqref{eq:diff_powers} can be considered negligible. The next part will consist, among others, of assessing if this assumption is reasonable. For that purpose, the two sets of equations (\ref{eq:pf_lin}) and (\ref{eq:p_approx}) will be compared to the exact powers.

\section{SIMULATIONS}
\subsection{Setup}
The method is tested on the IEEE 33-bus network \cite{ieee33bus}, an areal distribution grid comprised of 33 buses with 32 branches including 3 laterals.

\begin{figure}[h!]
    \centering
    \includegraphics[width=1\columnwidth]{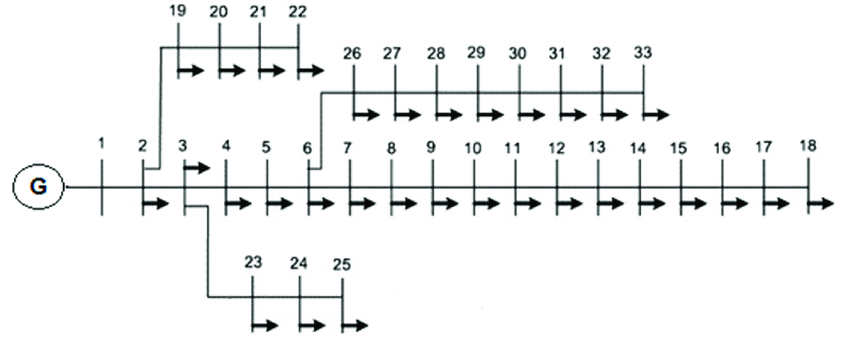}
    \caption{IEEE 33-bus distribution network.}
    \label{fig:ieee33}
\end{figure}

The simulations are run for a duration of one day, sampled every minute. The load profiles for each node are generated with $20\%$ Gaussian standard deviation variations around the nominal value. $0.01\%$ to $10\%$ and $10^{-4}\%$ to $0.1\%$ of Gaussian noise is added to the power and voltage magnitude measurements, respectively. In what follows, a noise level refers to the amount of Gaussian noise on the powers (e.g. 1\% means 1\% noise on $P$ and $Q$, and 0.01\% on $V$).

\subsection{Numerical implementation}
To perform the admittance matrix estimation, we use the algorithm from \cite{eiv}. The voltage being approximated by its magnitude, the covariance matrix (\ref{eq:cart_noise}) simplifies to a diagonal matrix, where its block matrices are:
\begin{subequations}
\label{eq:cart_noise_simp}
\begin{align}
    & \text{Var}[\Delta c | \tilde{v}, \tilde{\theta}=0] = \sigma_\epsilon, \\
    & \text{Var}[\Delta d | \tilde{v}, \tilde{\theta}=0] = \sigma_\delta \tilde{v}^2, \\
    & \text{Cov}[\Delta c, \Delta d | \tilde{v}, \tilde{\theta}=0] = 0,
\end{align}
\end{subequations}
where $\tilde{v} e^{j\tilde{\theta}} = (c + \Delta c) + j(d + \Delta d) = \tilde{c} + j \tilde{d}$. In the identification procedure, the standard deviation of the imaginary part $\sigma_\delta$ of the voltage is set to a large value of $100\sigma_\delta$, to mitigate the impact of the phase being unknown. As a comparison, we also use the adaptive Lasso method from \cite{tomlin}.


\subsection{Error metrics}
The Relative Root Mean Square Error (RRMSE) is used as the metric to compare the results in different scenarios.
\begin{equation}
\label{eq:rrmse_x}
    \text{RRMSE} = \frac{|| X_\mathrm{exact} - X ||_F}{||X_\mathrm{exact}||_F},
\end{equation}
where $X$ is the admittance matrix, or the active or reactive power matrices for all nodes and samples in time, either from (\ref{eq:pf_lin}) or from (\ref{eq:p_approx}). Note that this metric is calculated using the exact values of the admittance matrix and the voltages. To have an idea of the absolute error, we will also use the Mean Absolute Deviations (MAD): $\mathrm{MAD}_x = E[|x - x_{\mathrm{exact}}|]$.



\subsection{Results}

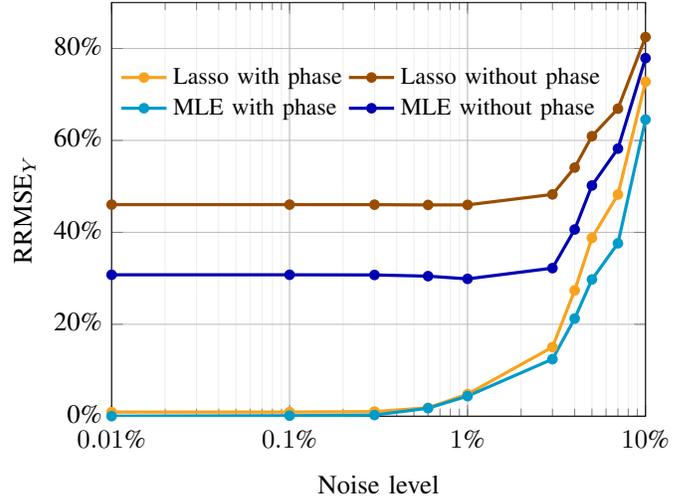
\begin{figure}[h!]
\centering
\begin{tikzpicture}
    \begin{axis}[
    xmode=log,
    xmin = 1e-4, xmax = 1e-1, ymin = 0, ymax = 90,
    grid = both,
    major grid style = {lightgray},
    minor grid style = {lightgray!25},
    width = 0.98\columnwidth, height = 0.8\columnwidth,
    xlabel = {Noise level}, ylabel = {$\text{RRMSE}_Y$},
    ytick = {0, 20, 40, 60, 80},
    yticklabels={0\%, 20\%, 40\%, 60\%, 80\%},
    xtick={0.0001, 0.001, 0.01, 0.1},
    xticklabels={$0.01$\%, $0.1$\%, $1$\%, $10$\%},
    tick label style={font=\normalsize},
    xticklabel style={name=tick no \ticknum},
    typeset ticklabels with strut,
    font=\normalsize,
    legend style={
        legend cell align=left,
        at={(0.47,0.87)},
        anchor=north,
        font=\small,
        legend columns=2,
        fill=none,
        draw=none,
    }
    ]
\addplot[very thick,YellowOrange, mark=*, mark options={scale=0.7}] table[meta=y]{y_data_ieee33/y_lasso_phase.txt};
\addplot[very thick,orange!60!black,mark=*, mark options={scale=0.7}] table[meta=y]{y_data_ieee33/y_lasso_nophase.txt};
\addplot[very thick,cyan!80!black,mark=*, mark options={scale=0.7}] table[meta=y]{y_data_ieee33/y_mle_phase.txt};
\addplot[very thick,blue!70!black,mark=*, mark options={scale=0.7}] table[meta=y]{y_data_ieee33/y_mle_nophase.txt};
\legend{
    Lasso with phase, 
    Lasso without phase,
    MLE with phase,
    MLE without phase,
}
    \end{axis}
\end{tikzpicture}
\caption{Comparison of the methods for various noise levels.}
\label{fig:noiselevel}
\end{figure}

In Fig. \ref{fig:noiselevel}, we observe that missing phase measurements induce about $30\%$ additional error, independently of the noise level. To qualitatively understand the influence of adding the current and voltage approximations, the Tables \ref{tab:results} and \ref{tab:results_i} provide better insights. Note that these metrics are almost independent of the noise level so they are computed for a noise level of $0.1\%$.

\begin{table}[h!]
\caption{RRMSE and MAD of the power flow equations and the MLE approximated relative to the exact powers. 
}
\label{tab:results}
\resizebox{\columnwidth}{!}{%
\begin{tabular}{l|cc|l}
& \multicolumn{2}{c|}{RRMSE {[}\%{]}} & MAD \\ \cline{2-3}
& \multicolumn{1}{l|}{Power Flow equations} & \multicolumn{1}{l|}{Adapted MLE problem} & [MW] \\ \hline
\multicolumn{1}{l|}{Active power P}   & \multicolumn{1}{c|}{1.39}   & 1.33  & \multicolumn{1}{c}{0.0031} \\ \hline
\multicolumn{1}{l|}{Reactive power Q} & \multicolumn{1}{c|}{8.64} & 6.19  & \multicolumn{1}{c}{0.0022} \\
\end{tabular}%
}
\end{table}

The results (Table \ref{tab:results}) point out that the approximation of the MLE problem has power matrices relatively close to those of the linearised power flow equations. Reminding that the power flow equations are also approximated considering the small angles, the MLE approximation can be better than the power flow equations. This shows that if the power flows can be linearised, then the term of difference between the two approximations can be neglected, as intuitively assumed in \hyperlink{subsectionVB2}{part V.B.2}.

\begin{table}[h!]
\caption{RRMSE of the current approximation relative to the exact current.}
\label{tab:results_i}
\centering
\begin{tabular}{l|c}
     & \multicolumn{1}{l}{RRMSE [\%]} \\ \hline
$\text{RRMSE}_{\Re(\underline{I})}$ & 0.28 \% \\ \hline
$\text{RRMSE}_{\Im(\underline{I})}$ & 6.24 \%
\end{tabular}
\end{table}

The approximation of the real part of the current is close to its exact value whereas the imaginary part deviate more. This is coherent with \hyperlink{app2}{assumption 2} because the phase influences almost exclusively the imaginary part of $\underline{I}$.

\begin{figure}[h!]
\begin{subfigure}[t]{0.24\textwidth}
    \centering
    \includegraphics[width=1\textwidth]{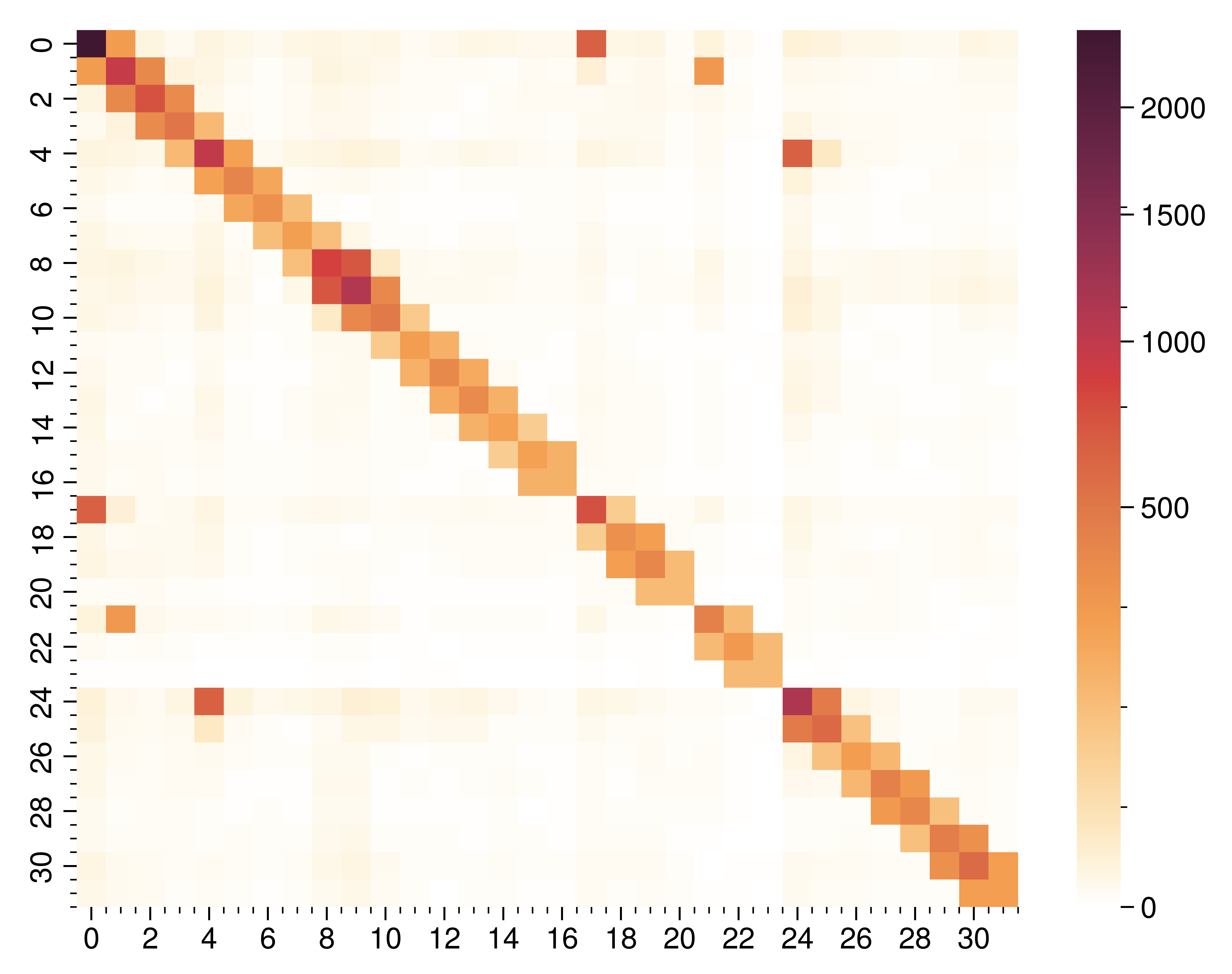}
    \caption{Admittance matrix with the phase.}
    \label{ig:y_with_p}
\end{subfigure}
\hfill
\begin{subfigure}[t]{0.24\textwidth}
    \centering
    \includegraphics[width=1\textwidth]{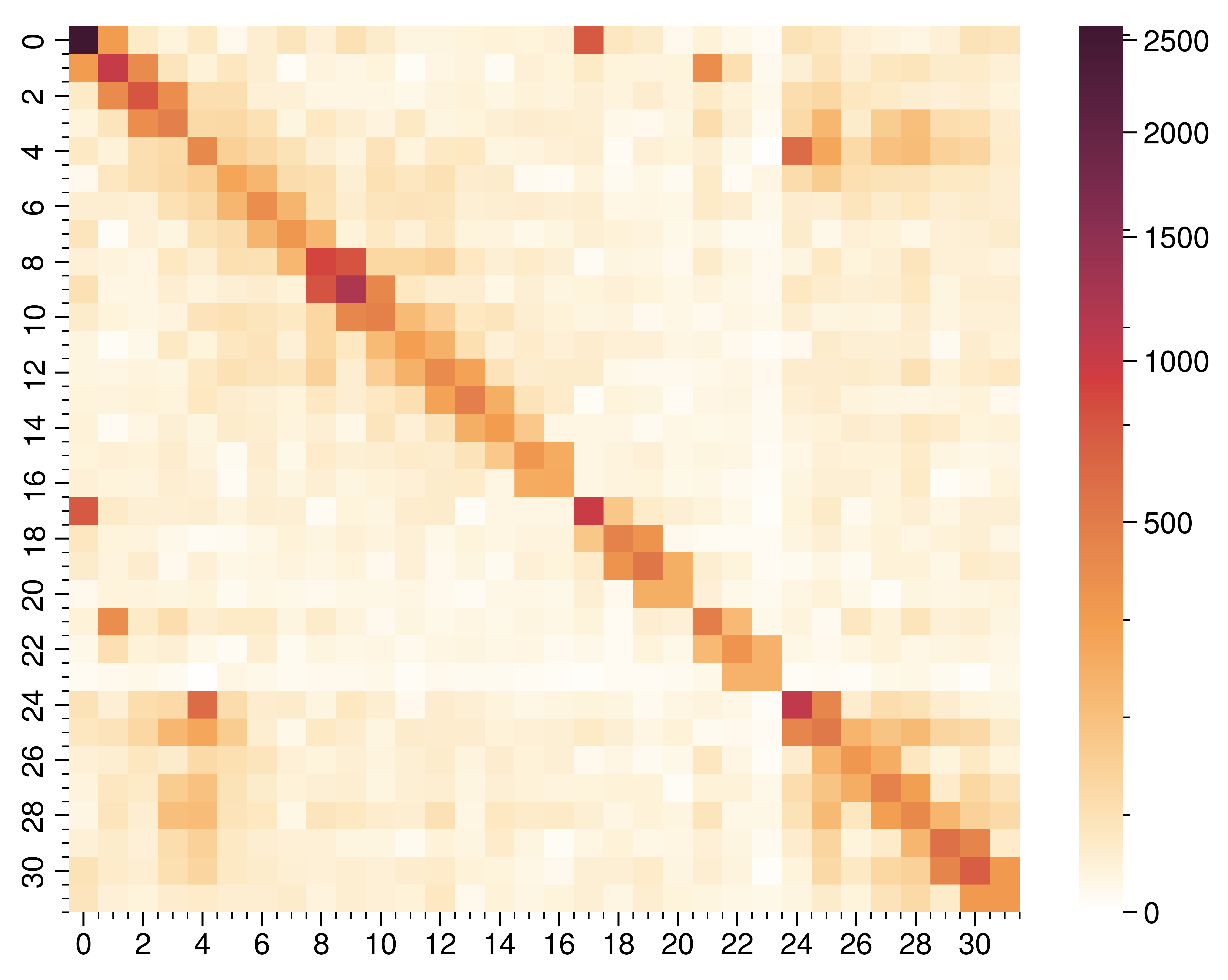}
    \caption{Admittance matrix without the phase.}
    \label{ig:y_no_p}
\end{subfigure}
\caption{Admittance matrices with or without the phase. The admittance matrices are inspected for a noise level of $1\%$. Colors are in log scale.}
\label{fig:y}
\end{figure}

Regarding the admittance matrices (Fig. \ref{fig:y}), numerous zero elements in the lower and upper triangle matrices are not considered as such in the absence of the phase. Although most of them have small magnitudes, the sparsity of the admittance matrix is lost. Adding a penalty term in the objective function could prevent such behaviour. The method \cite{eiv} does propose Maximum A Posteriori estimates, based on the Bayes' rule, which add a $\ell1$-norm to the MLE problem. 


\section{DISCUSSION}
As previously noticed, when the phase is not measured, the RRMSE of the admittance matrices is shifted above by around $30\%$ even for small noise level (see Fig. \ref{fig:noiselevel}). A plausible reasoning comes from the initial equation (\ref{eq:cv}). By expressing the current with a phase-shift with respect to the voltage: $\underline{v}_h = v_h e^{j \alpha_h}$ and $\underline{i}_h = i_h e^{j (\alpha_h + \varphi_h)}$, a diagonal matrix composed of the angles $\bm{\alpha}$ can be defined: $D = \text{diag} (e^{j\bm{\alpha}})$, such that $D^*D=\mathcal{I}_n$. Equation (\ref{eq:cv}) reads as:
\begin{equation}
    D \underline{\bm{i}} = D \underline{Y} D^* D \underline{\bm{v}}.
\end{equation}
Therefore, when the phase is unknown, one must use the approximation $(\underline{s}^* \oslash v) = \underline{A} v$, where $\underline{A}$ should be the admittance matrix $\underline{Y}$. In practice, one is actually identifying the time-varying matrix $ \underline{A} = D\underline{Y}D^*$.

All information relative to the phasors synchronisation is lost, meaning that it is impossible to seize changes of phase angles between nodes. Reducing this error requires to minimize the voltage phase variations while maximizing the voltage magnitude ones. This can be done by specific load profiles or low X/R ratios but in practice both cannot be easily influenced.

\section{CONCLUSION}
This paper introduces non-PMUs measurements in the method \cite{eiv} to identify the admittance matrix of distribution grids. The MLE problem formulated for PMUs measurement devices was adjusted accordingly by considering two assumptions. They resulted in a greater error of $30\%$ in the admittance matrix. Besides, the admittance matrix has lost its initial sparsity. Finally, a comparison in terms of powers proved that the MLE constraint is even more accurate than the power flow equations. 

Future improvements could consist in approximating the voltage phase to a different value than zero, for instance to its average value over a year, if such data is available. The method could also be further implemented to combine buses equipped either with $\mu$PMUs or with Smart Meters in order to deal with grids having both types of measurements.





\printbibliography[title=References,heading=bibintoc]

\end{document}